# Molecular Beam Epitaxy Grown Cr$_2$Te$_3$ Thin Films with Tunable Curie Temperatures for Spintronic Devices

*Hongxi Li[#,†,‡,||], Linjing Wang[#,†], Junshu Chen[#,†,¶], Tao Yu[†], Liang Zhou[†], Yang Qiu,[†]*

*Hongtao He[†], Fei Ye[†], Iam Keong Sou[∗,‡], and Gan Wang[∗,†,§]*

† Institute for Quantum Science and Engineering, and Department of Physics, Southern University of Science and Technology, 1088 Xueyuan Avenue, Shenzhen 518055, China

‡ Department of Physics, The Hong Kong University of Science and Technology, Hong Kong 999077, China

¶ Department of Physics, National University of Singapore, 2 Science Drive 3, Singapore 117551, Singapore

§ Shenzhen Key Laboratory of Quantum Science and Engineering, 1088 Xueyuan Avenue, Shenzhen 518055, China

|| Current address: Department of Physics and Astronomy, 525 Northwestern Avenue, West Lafayette, IN 47907, United States

# These authors contributed equally.

* Emails of corresponding authors: *phiksou@ust.hk; *wangg@sustech.edu.cn

## ABSTRACT

Materials with perpendicular magnetic anisotropy (PMA) effect with high Curie temperature is essential in spintronics applications. $Cr_2Te_3$ is a transition metal chalcogenide that demonstrates PMA effect but with a relatively low Curie temperature of about 180 K, significantly limiting its practical application. In this work, we reported the epitaxial growth of $Cr_2Te_3$ thin films on $Al_2O_3$ substrates with Curie temperature ranging from 165 K to 295 K, closely dependent on the thicknesses and lattice constants of the thin films. To study the physical origin of the improved Curie temperature, structural analysis, magneto-transport and magnetic characterizations were conducted and analyzed in detail. In contrast with previous reports, all the high Curie temperature thin films have electron type carriers instead of hole carriers, which is possibly caused by the Al diffusion from the $Al_2O_3$ substrate. Based on the structure and chemical analysis, a phenomenological model based on the degree of coupling between ferromagnetic and anti-ferromagnetic ordering, hosted by the fully-occupied and with-vacancy Cr layer alternatively, was proposed to explain the observed Currie temperature enhancement in our samples. These findings indicate that the Curie temperature of $Cr_2Te_3$ thin films may potentially be tuned by Al doping, performing as a novel magnetic material suitable for various magnetic applications.

**KEYWORDS**

# INTRODUCTION

Being a typical transition metal chalcogenide, chromium telluride possesses a large family of compounds in the bulk state. All these compounds are of Hexagonal structure and ferromagnetic ordering at stable states with a wide range of Curie temperature from 170 K to 340 K, depending sensitively on the atomic ratio between Cr and Te. In the bulk state, the chromium telluride with Cr/Te ratio around 1:1 has the highest Curie temperature around 340 K,[1-3] while the other phases of chromium telluride with Cr vacancy around 0.1 to 0.33 have Curie temperature between 150 K and 340K. Not only the Curie temperature, the magnetic anisotropy of the chromium telluride compounds also depends on the stoichiometric ratio as well as the lattice structure intimately, in which only the phase with $Cr_2Te_3$ has stable perpendicular magnetic anisotropy in thin film form[4-5]. In the past decade, many efforts had been made on the growth of chromium telluride thin films on semiconducting substrates. For example, structural and magnetic properties of hexagonal $Cr_2Te_3$ films grown on CdTe(001) with Curie temperature of 175 K has been studied by Ken Kanazawa et al.[6] A perpendicular magnetic anisotropy (PMA) and spin-glass-like behavior of hexagonal $Cr_2Te_3$ grown on $Al_2O_3$(0001) and Si(111) substrates with Curie temperature of about 180 K were studied systematically.[7-8] The thin $Cr_2Te_3$ ferromagnetic metallic film was used in a field effect capacitor (FEC) structure at temperature of 175 K.[9] Saito et al. have studied tunneling magnetoresistance (TMR) in the magnetic tunneling junctions (MTJs) with $Cr_{1-\delta}Te$ being one of the electrodes around 170 K.[10] Especially in the recent research, the $Cr_2Te_3$'s strong PMA has been widely integrated in the interface engineering for realizing non trivial topological spin textures, such as the magnetic skyrmions[11-13], encouragingly showing that the hexagonal phase $Cr_{1-\delta}Te$ thin films possess promising potentials for the application in the spintronic and spin transfer torque (STT) device development. However, the Curie temperature of the epitaxially grown chromium telluride thin films with strong PMA is still desired to be further improved for fabricating reliable devices working at room temperature.

Even though the $T_C$ for bulk hexagonal phase $Cr_{1-\delta}Te$ can reach as high as 340 K when Cr/Te ratio reaches 1:1, their easy axis is orthogonal to the c-axis, which made it unlikely to be grown with strong PMA in devices.[1, 14] Several works have been reported for the growth of Zinc blende CrTe on GaAs (001) substrate with ZnTe or CdTe buffer layer. [9, 15-17] However, the fragile nature of the zinc blend structure strongly prevents its practical application. Hereby, we report that $Cr_2Te_3$ thin films with a stable NiAs-type hexagonal structure grown on $Al_2O_3$ (0001) substrates by the MBE technique with high Curie temperature. Thorough studies on the strain and lattice constant show that the Curie temperature is sensitively determined by the (001) plane spacing of $Cr_2Te_3$ thin films. To the best of our knowledge, our work achieved the highest Curie temperature for NiAs-type chromium telluride thin films grown by epitaxial schemes, which paves the way for the fabrication of chalcogenide based spintronic devices working at near ambient conditions.

**EXPERIMENTAL DETAILS**

[001] oriented $Cr_2Te_3$ thin films were grown on epi-ready $Al_2O_3(0001)$ substrates (HF-Kejing, China) in a customized MBE system with base ultrahigh vacuum better than $1 \times 10^{-10}$ Torr, equipped with a 15 keV reflective high energy electron diffraction (RHEED) for monitoring the epitaxy. Before the growth, $Al_2O_3(0001)$ substrates were annealed at 600 °C for 15 minutes to degas. Chromium (99.999% Goodfellow UK) and tellurium (99.999% Goodfellow UK) were evaporated by two Knudsen-type effusion cells for providing independent fluxes, respectively, and co-deposited on the substrate at 370°C, as shown in Scheme 1. The flux ratio between chromium and tellurium atom was determined to be around 1:7. Ex-situ thickness characterization is done by atomic force microscopy (AFM) and X-ray Reflectometry (XRR), indicating that the growth rate was about 1nm/min. To investigate the property of $Cr_2Te_3$ grown on $Al_2O_3$ substrates systematically, several $Cr_2Te_3$ samples with different thicknesses named as sample #1 (15 nm), #2 (30 nm), #3 (45 nm) and #4 (300 nm) are reported in this work.

A Rigaku High-resolution X-ray diffractometer (XRD), Talos transmission electron microscope (TEM) coupled with energy dispersive x-ray spectrometer (EDX) and FEI Titan Spherical aberration corrected transmission electron microscope (STEM) were employed to study the lattice structure and atomic ratio of the as-grown chromium telluride thin films (the detail stoichiometry analysis of $Cr_2Te_3$ can be found in section S1 in the supporting material). The magnetic properties were studied using Quantum

Design superconducting quantum interference diffractometer (SQUID). Transport properties are measured using an Oxford Physical Property Measurement System (PPMS).

**Scheme 1.** Schematics of high Curie temperature $Cr_2Te_3$ grow on $Al_2O_3$ by Molecular Beam Epitaxy (MBE).

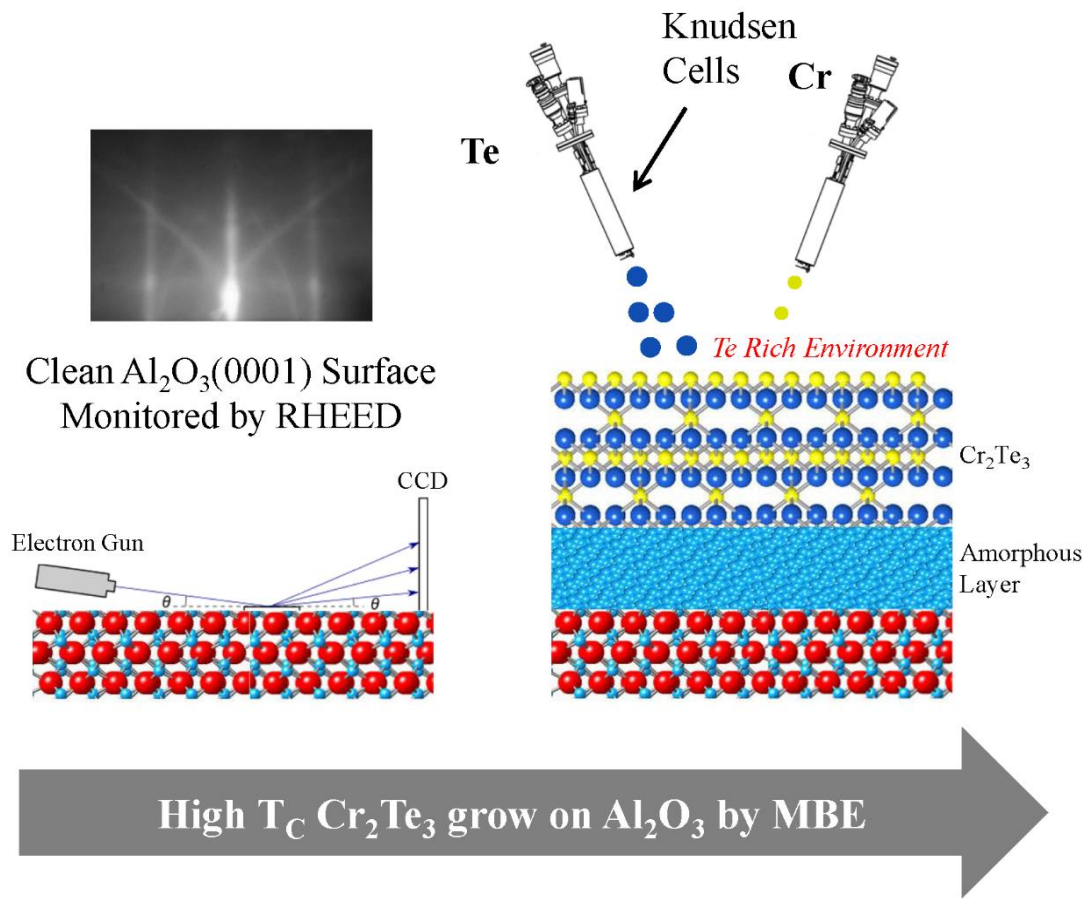

Figure 1 displays the RHEED patterns observed for the growth of a typical $Cr_2Te_3$ thin film sample. Figure 1(a) and (b) are the RHEED patterns of a cleaned $Al_2O_3(0001)$ surface along the [10$\bar{1}$0] and [11$\bar{2}$0] electron beam incidence, respectively, after a pre-heat treatment. Kikuchi lines can be clearly seen in these patterns, indicating a clean and smooth surface morphology of the substrate was achieved. Corresponding RHEED patterns observed after the growth of a Cr$_2$Te$_3$ thin film are shown in Figure 1(c) and (d). Half-order reconstruction lines are seen in both patterns, reflecting that the as-grown Cr$_2$Te$_3$ thin film has a high crystalline quality. The long streaky patterns were being maintained during the entire Cr$_2$Te$_3$ growth process, suggesting that its growth was via a two-dimensional growth mode with an atomically flat surface.

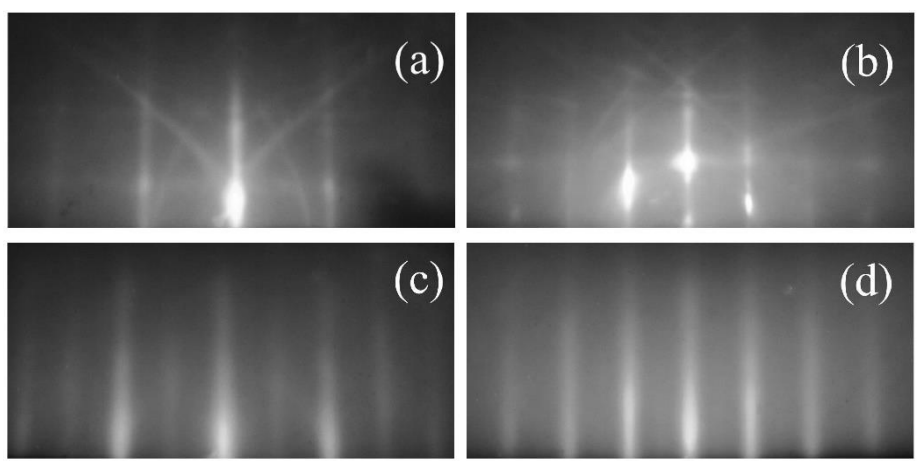

**Figure 1.** RHEED patterns observed for the growth of a typical Cr$_2$Te$_3$ thin film. (a) & (b) RHEED patterns from a clean Al$_2$O$_3$ (0001) surface with the incident electron beam along the [10$\bar{1}$0] and [11$\bar{2}$0] orientations of Al$_2$O$_3$, respectively; (c) & (d) Corresponding RHEED patterns from the Cr$_2$Te$_3$ surface along the same orientations.

Figure 2(a) shows the XRD profile of sample #2, showing a typical result for the Cr$_2$Te$_3$ thin film samples presented in this work. The layer peaks (004) and (008) in Figure 2(a) match well with the hexagonal structure of Cr$_2$Te$_3$ oriented along the [001] direction, while the (006) peak is from the Al$_2$O$_3$ substrate. The derived c-lattice parameter is 12.985 Å for Al$_2$O$_3$, same as the standardized database in Springer Materials.[18] The c-lattice parameter of the Cr$_2$Te$_3$ layer of sample #2 is 12.304 Å. The thickness dependence of the c-lattice parameter of the layer peak was further studied among the four Cr$_2$Te$_3$ samples. Figure 2(b) shows the XRD (008) peaks of these samples (samples #1 ~ #4) with their c-lattice parameters determined to be 12.255 Å, 12.304 Å, 12.309 Å, and 12.327 Å, indicating that the c-lattice parameter increases with sample thickness. It is worth to note that all these values are larger than the reported c-lattice parameter of bulk hexagonal Cr$_2$Te$_3$, which is 12.07 Å.[19] We believe that this difference may be caused by unexpected Al incorporation in the layers through diffusion from the Al$_2$O$_3$ substrate, which will be addressed later.

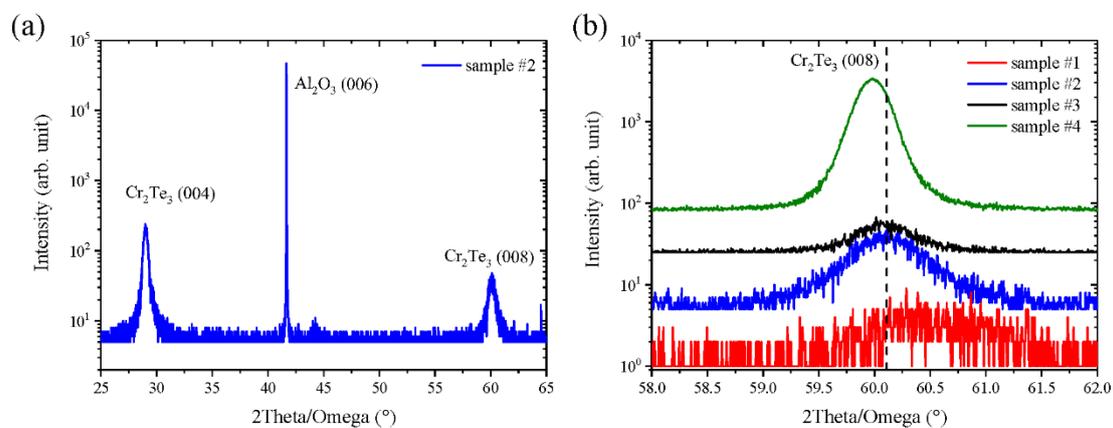

**Figure 2.** X-ray diffraction (XRD) profiles of Cr$_2$Te$_3$ thin films. (a) XRD profile of sample #2; (b) XRD profiles of the (008) peak of Cr$_2$Te$_3$ thin films with increasing thickness.

The magnetization versus temperature (M-T) curve from 2 K to 320 K under an applied field of H = 1000 Oe perpendicular to the surface and resistance versus temperature (R-T) curve from 2 K to 300 K for sample #2 are shown in Figure 3(a). Both the M-T and R-T curves show a typical behavior of ferromagnetism around their para-ferro transition upon cooling. Curie temperature is obtained by performing the first derivation of these curves. As shown in Figure 3(b), the peak and valley at 244 K in the derivative curves indicate the Curie temperature at which magnetization and resistance change abruptly. We have carried out similar studies for the other samples. Figure 3(c) displays the M-T curves for all four samples, in which, sample #4 was under a magnetic field of 50 Oe to show a clearer low-temperature behavior, which may due to the strong magnetic anisotropy. These M-T curves show that a tuning of Curie temperature from 165 K to 295 K can be achieved by varying their thicknesses that somehow might be related to the c-lattice parameter as illustrated in Figure 2(b). In the following paragraphs, we will present the results of further structural, magnetic and magneto-transport measurements to address the underlying mechanism of the observed tunability of the Curie temperature of our $Cr_2Te_3$ samples.

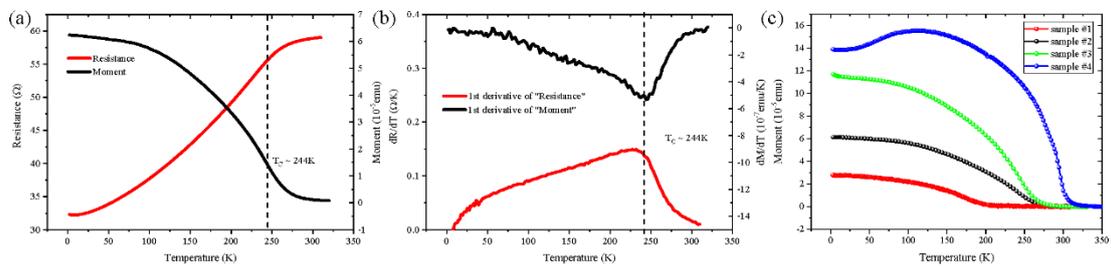

**Figure 3.** The magnetization characterization of sample #1~#4. (a) R-T and M-T curves of sample #2; (b) First derivation of R-T and M-T curves of sample #2; (c) M-T curves

measured for samples #1~#4 with magnetic field perpendicular to the surface, in which the applied field for sample #1~#3 is 1000 Oe, except 50 Oe for sample #4

Figure 4 displays the typical cross-sectional HAADF STEM images near the interface between $Cr_2Te_3$ and $Al_2O_3$ viewing along the $[11\bar{2}0]$ direction of $Al_2O_3$, cutted from a $Cr_2Te_3$ sample with a same thickness of sample #2, while a ZnSe layer was capped for protection purpose. Two types of domains with different orientations of $Cr_2Te_3$, one along the $[110]$ direction (Figure 4(a)) and the other along the $[\bar{1}\bar{2}0]$ direction (Figure 4(c)) are found. As estimated from all the available TEM images, we found that more domains that atoms arranged like Figure 4(c) were found than those arranged like Figure 4(a). Structural simulations for both STEM images are given accordingly in Figure 4(b) and 4(d). It should be noted that an amorphous layer with thickness ~ 2 nm has been observed at the interface between the $Cr_2Te_3$ layer and the $Al_2O_3$ substrate for both domain types. Chemical composition characterization of the $Cr_2Te_3$ layer in the area that is far away from the interface is performed by EDX installed in the STEM. Figure 4(e) displays the STEM image from a domain like the one shown in Figure 4(c), which is used for this characterization. The resulted mapping for Cr, Te and Al are shown in Figure 4(f), (g) and (h) respectively. The distribution of Te element shown in Figure 4(g) is in good consistency with the STEM graph shown in Figure 4(e). Cr atoms which is invisible in the TEM graph due to its small atomic radius, can be observed in the EDX mapping. The EDX mapping result of Cr atoms (Figure 4(f)) shows that Cr atoms represented by relative bright blue dots are interspersed between Te atoms, consistent with the structure as shown in Figure 4(d). Besides Cr and Te elements, small amount of Al is identified in Figure 4(h).

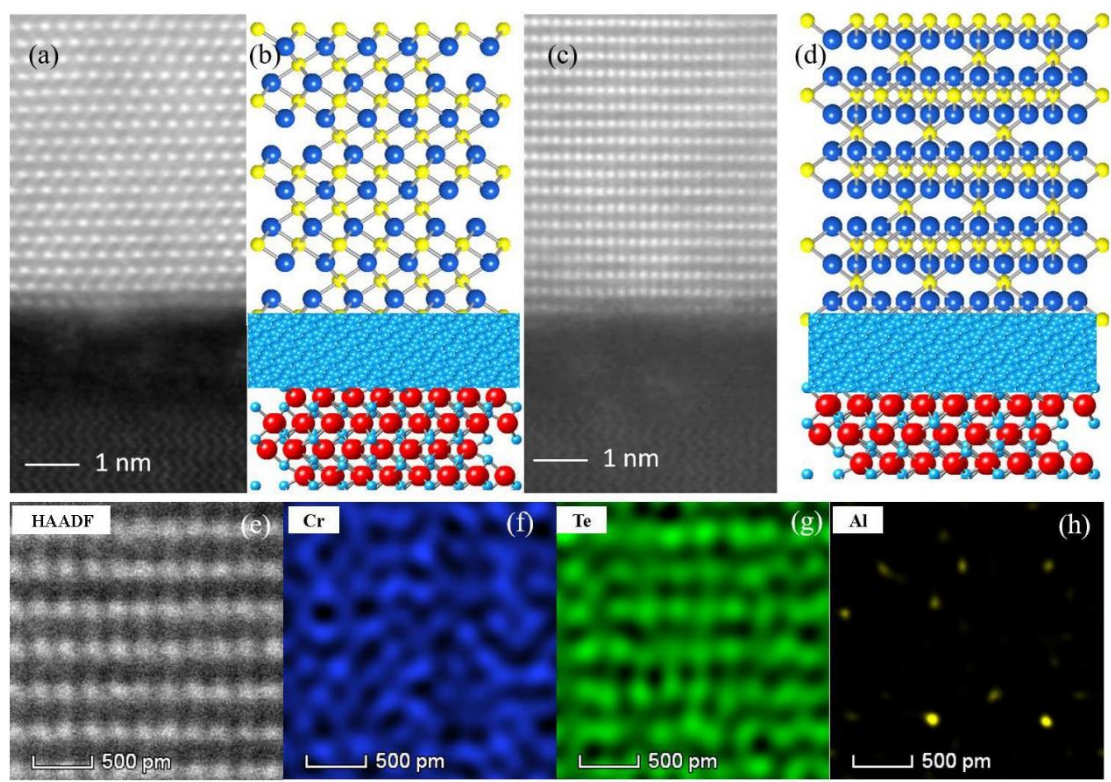

**Figure 4.** High angle annular dark field (HAADF) scanning transmission electron

microscopy (STEM) images and EDX mapping results of a $Cr_2Te_3$ sample. (a) & (c)

The cross-section TEM graph of the interface, (b) & (d) show the structure simulation

corresponding to (a) & (c) respectively (e): an area in (c) used for EDX mapping. (f) &

(g) & (h) show EDX mapping of Cr, Te, and Al element respectively for the area shown

in (e).

To determine the composition of the amorphous layer shown in Figure 4, we performed

atomic-scale EDX mapping on the interface region. The STEM graph taken in HAADF

mode is shown in Figure 5(a), showing the atomic structure and elements distribution of the interface. In Figure 5(a), we scan lines across the interface (amorphous layer) between the $Cr_2Te_3$ layer and the $Al_2O_3$ substrate. The corresponding distribution results of the Cr, Te, Al and O atomic fraction along the line is shown in Figure 5(b), with the HAADF intensity also added for assisting the chemical analysis, in which the oscillations of Te and Cr atomic fraction curve dropped abruptly at the interface of the amorphous layer, followed with a rising chemical composition of Al and O, manifesting a sharp interface between $Cr_2Te_3$ and the amorphous layer. Surprisingly, we observe a significant Al rich (with the ratio nearly Al:O $\approx$ 2:1) in the amorphous layer, indicating that the amorphous layer is likely a Al dominated layer and the thickness is about 2 nm. Similar observation for the growth of GaN on $Al_2O_3$ has been reported by Wenliang Wang et al.[20] They named it as nitridation effect of $Al_2O_3$. To investigate the formation mechanism of the unknown Al dominated amorphous layer, we further performed a control experiment by growing the $Cr_2Te_3$ at a lower temperature around 300 K. The structure analysis was shown in supporting materials (Figure S1), in which the amorphous layer does not exist at the interface (~300 °C). Hence, in our $Cr_2Te_3$ sample, we speculate that the tellurium-rich environment and high substrate temperature during the sample growth may play important roles for causing the top few layers of the $Al_2O_3$ substrate to suffer from oxygen deficiency.

Noticeably, the "noise" of the O element in Figure 5(a) is quite large in $Cr_2Te_3$ layer. However, it could not proof the O element diffuse into the top layer, there are two reasons may illustrate the existence of the O element at the top part: First, there is still a little "noise" exists due to the Al$_2$O$_3$ substrate background. Second, commonly, there exists lots of contamination of the O element from the air, thus the O element should be ubiquitous on the surface of TEM sample. However, we still could estimate that O element is not incorporated into the Cr$_2$Te$_3$ lattice from the EDS results shown in Fig 5 (b), as we could clearly observe that the O element curve is nearly uniform and the O element randomly distributes at Cr$_2$Te$_3$ layer as shown in Figure 5(b), it may suggest the O element is coming from the contaminations, while the Al curve shows a decline, which is a typical shape for diffusion. In contrast, as to the Al signal at the top region, we still could observe the Al elemental signal in Figure 4(h), where is relatively far away from the interface of Cr$_2$Te$_3$ and Al$_2$O$_3$, since the environment contains no Al element, hereby, we could be sure that the Al element is diffused into the Cr$_2$Te$_3$ lattice. Hereby, in this report, we believe the influence of the O element to Cr$_2$Te$_3$ should be negligible, since the sample is exposed to the air for measurements, and its properties do not vary within a considerable long time. Thus, we focus the main concern here on the Al incorporation caused by the diffusion effect.

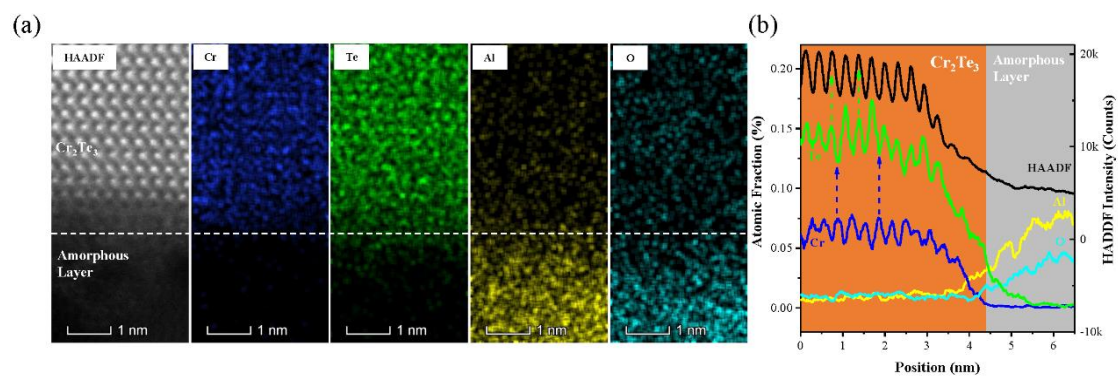

**Figure 5.** (HAADF) STEM images and EDX mapping analysis of the interface between $Cr_2Te_3$ and the amorphous layer. (a) $Cr_2Te_3$/$Al_2O_3$ interface used for EDX mapping and its corresponding EDX mapping of Cr, Te, Al and O element, respectively. (b) EDX mapping profile of (a), showing the relation of HAADF intensity and atomic fraction of Cr, Te, Al and O elements versus the vertical position of interface. The x-axis is the distance from the starting points of the image bottom, the approximate positions of the amorphous layer could be distinguished by the sudden drop of Te elements. Noticeably, the peaks of the Te match with the peaks of HAADF intensity curve as the green arrows indicated. Moreover, the peak of Cr located exactly at the valley of Te as the blue arrow indicated, demonstrating the Cr-Te-Cr layered structure of $Cr_2Te_3$, the excess Al atomic fraction in amorphous layer suggesting O deficient phase from $Al_2O_3$.

Now let us revisit the observed Al contamination in our $Cr_2Te_3$ layers and see how it may affect the electrical transport of these layers. To our best knowledge, the conduction type of all $Cr_2Te_3$ bulk or thin film materials reported previously is p-type.[5] $Cr_2Te_3$ samples grown on ZnSe in our group using the same MBE system also constantly show p-type feature.[21] To investigate the effect of Al contamination, A six-terminal Hall bar device was fabricated on a small piece of sample #2 for performing magneto-transport measurements. The width and length of the device are 1000 μm and 500 μm respectively. Its Hall resistivity was derived from the voltage along the XY direction with a current of 50 μA flowing along the XX direction. The results of Hall effect measurements performed on sample #2 are shown in Figure 6 with $\rho_{yx}(H)$ curves having been offset vertically for clarity. At 280 K, the measured Hall resistivity is linearly dependent on H field, which is a typical feature of ordinary Hall effect (OHE). However, below Curie temperature (244 K), a clear hysteresis behavior is observed, which is an indicator of the anomalous Hall effect (AHE) that arises from the ferromagnetic ordering with PMA property. At low temperature, the hysteresis loop narrows and vanishes at 100 K. As temperature decreases further, $\rho_{yx}$ reverses its sign comparing with $\rho_{yx}$ measured at temperature above 100 K. The observed temperature dependent AHE is related to several transport mechanisms of the charge carriers,[22] there are several mechanisms including the extrinsic (skew scattering, side jump) and intrinsic (Berry phase) that could cause the AHE.[23] Those mechanisms will compete with each other, leading to the 'sign-reversal' feature observed in Figure 6 (the details of our analysis will be reported elsewhere). Here, we would like to focus on the observation that the overall Hall resistance show decreasing tendency with increasing magnetic field, as shown in Figure 6(b), which indicates that sample #2 have n-type carriers. The carrier concentration is derived by linearly fitting the Hall resistance $\rho_{yx}$ to magnetic field at large magnetic field values (ordinary hall effect dominating region), which is estimated to be about $10^{21}$ cm$^{-3}$ for sample #2. We also performed Hall measurement of the control sample shown in supporting materials (a $Cr_2Te_3$ thin film grown on $Al_2O_3$ substrate without Al rich amorphous layer), in which p-type hall conductance was shown unambiguously (Figure S2). All above results lead us to believe that the observed Al -rich interface layer and incorporation of Al atoms in $Cr_2Te_3$ lattice may cause our $Cr_2Te_3$ thin films grown on $Al_2O_3$ substrate to be n-type, possibly by contributing free electrons from the Al atoms.

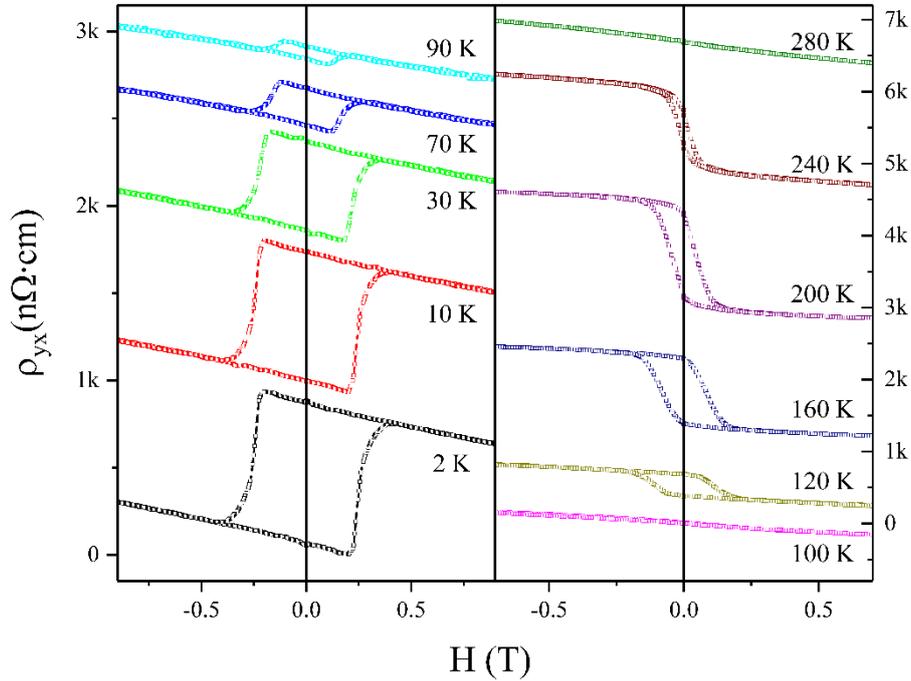

**Figure 6.** Results of Hall effect measurements performed on sample #2. Showing the temperature dependent $\rho_{yx}$ vs magnetic field strength from 2 K to 280 K of sample #2.

The magnetic hysteresis loop measured for sample #2 is shown in Figure 7. The measured magnetization curve (M-H curve) with the applied magnetic field perpendicular to the sample surface shows a square shaped loop with coercive force around 0.25 T and shifts into a round shape when the applied magnetic field is parallel to the sample surface, indicating that the sample has uniaxial anisotropy with easy axis perpendicular to the plane, consistent with previous reports on the easy axis of $Cr_2Te_3$ compound.[7] However, the hysteresis loop is not saturated with magnetic field up to 1.5 T, indicating the relatively large anisotropy of $Cr_2Te_3$ may be due to the heavy element

Te and low symmetry of the crystal lattice.[24-25] In Figure 7, a small step-like kink feature around the zero-field region is found in the hysteresis loop for out of plane direction, which is usually observed in complicate magnetic systems with composite ordering, such as ferrimagnetic or ferromagnetic materials inclusion with different switching fields. Noticeably, Fang Wang et al. have also observed such kink in $Cr_2Te_3$ nanorods, which could be ascribed to spatial inhomogeneity of the nanorods.[26] Similarly, via TEM images of Figure 4, we have demonstrated that some rotation domains exist in our $Cr_2Te_3$ thin film. Hereby, we believe that the grain boundary of $Cr_2Te_3$ rotation domain may attribute to the kink in MH loop, however, the details of the formation mechanism of the kink is still unrevealed, which is worthy for the further study.

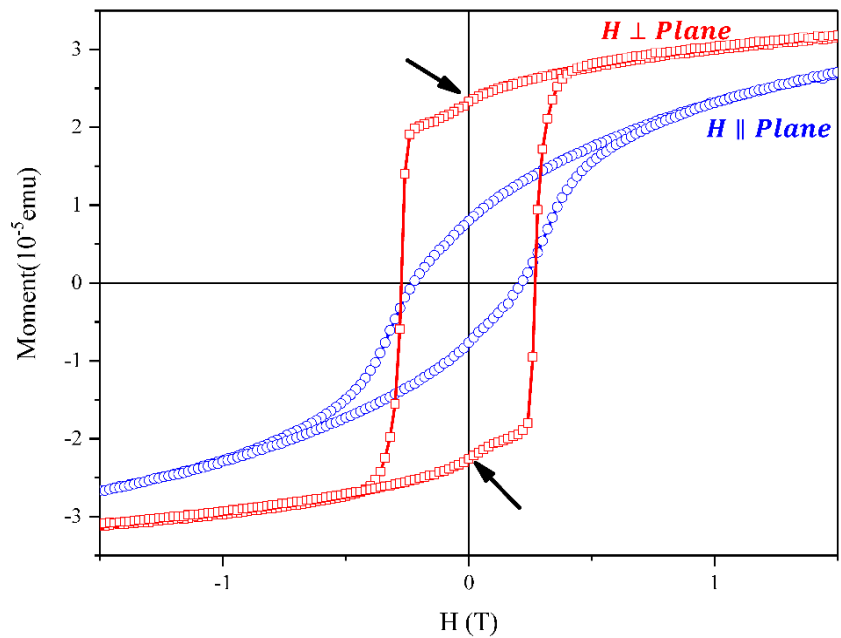

**Figure 7.** M-H curves of sample #2 with magnetic field perpendicular and parallel to the surface respectively measured at 2.1 K. The arrows indicate the small step-like kinks.

To study the magnetization properties of $Cr_2Te_3$ thin films, zero field cooled (ZFC) and field cooled (FC) M-T curves are measured simultaneously. Figure 8(a) shows the ZFC and FC curves of sample #2 with a magnetic field perpendicular to the sample surface with the magnitude of 100 Oe, 500 Oe, 1000 Oe and 2000 Oe, respectively. FC curves were measured while cooling down from 320 K. ZFC curves were measured while warming up with the same magnetic field used for measuring FC curves after cooling from 320 K without magnetic field. The obtained FC magnetization curves correspond to a typical magnetization behavior of ferromagnetism. The ZFC curves display an increase of magnetization when warming up from 2 K for all curves, followed by a decrease of magnetization after a characteristic temperature, and eventually merge with the FC magnetization curves at a blocking temperature $T_b$. With increasing magnetic field, the blocking temperature decreases. This temperature dependent magnetization behavior is similar to the typical behavior of spin-glass or spin-cluster. Noted that the spin-glass or spin-cluster should be distinguished by the decaying magnetic remanence moments after withdrawing external magnetic field.[27] To verify this feature, Magnetic remanence ($M_r$) of sample #2 was conducted at 2.1 K after applying a 2 T magnetic field perpendicular to the sample surface and withdrawing back to zero quickly(Figure 8(b)), in which the out of plane $M_r$ stays at a constant value with measuring time at least up to 60 mins, which is a signature of non-decay ferromagnetic ordering but not spin glass. Thus, we believed that the occurrence of discrepancy in ZFC and FC M-T curves of sample #2 are likely not a signature of pure spin-glass or spin cluster, hereby, we believe that the magnetization behavior in Figure 8(a) is due to antiferromagnetic/ferromagnetic coexistence in $Cr_2Te_3$. In fact, it is consistent with previous research on the chromium telluride system, which experimentally show that in the Cr fully occupied layer of chromium telluride, moments are ferromagnetically ordered,[5] while in the Cr vacancy layer, the moments are antiferromagnetically ordered.[2, 4, 28-29]

To investigate the details of the magnetic ordering in our samples, magnetic hysteresis loops of sample #1-#4 were shown in Figure 8(c), in which the magnetic moments are normalized to -1 to 1 and offset vertically for clarity. we could clearly observe the kink disappears and coercive field decreases with the thin film thickness increasing, in which the former feature could result from the reduction of rotation domain during the thin film growth process. Our thickness dependent structural analysis (Figure 2) shows that, with increasing thin film thickness, the distance between the inter-atomic layers stacking along [001] direction of $Cr_2Te_3$ gets larger, weakening the interaction between the Cr vacancy layer and the fully occupied layer. Considering the small AFM ordering exists in the Cr vacancy layer, with increasing distance between the atomic layers, the coupling between FM ordering and AFM ordering is expected to be weakened either. Thus, it is easier for moments ferromagnetically ordered to be aligned to the external magnetic field, which may serve as the mechanism for explaining the $H_C$ and Curie temperature dependence on the thickness in Figure 8(c) and Figure 3(c). It is worth noting that previous high-pressure studies on $Cr_2Te_3$ samples grown by others indicate that the Curie temperature decreases with the contraction of c-axis.[30-31] In our work, the observed enhancement of $T_C$ in our samples can be attributed to the same mechanism with an opposite direction that the c-axis is extended.

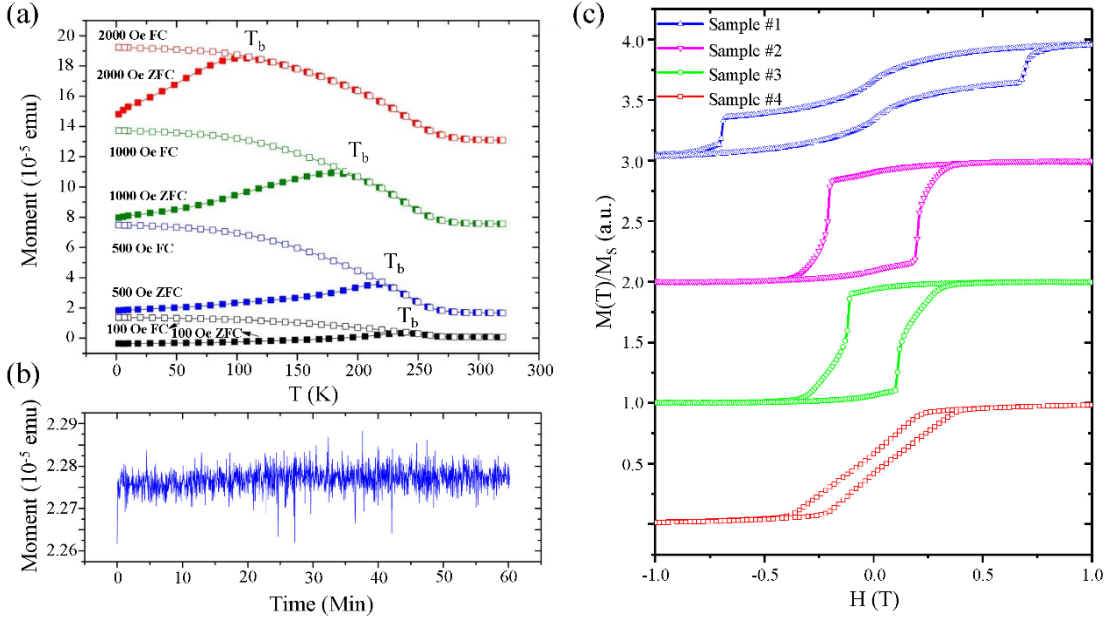

**Figure 8.** Results of various magnetic moment measurements performed on $Cr_2Te_3$ samples. (a) ZFC and FC curves of sample #2 with magnetic field perpendicular to the surface; (b) Magnetic remanence (Mr) of sample #2 at 2.1 K after applying a 2 T magnetic field perpendicular to the sample surface and withdrawing back to zero quickly; (c) Magnetic hysteresis loops measured for sample #1~#4 with magnetic field perpendicular to the surface at 2 K.

Combining the transport and magnetic measurement results, we observe an unusual thickness dependent c-axis variation in our n-type $Cr_2Te_3$ with higher Curie temperature, which is absent of reported in previous study on the p-type $Cr_2Te_3$ thin films grown on the ZnSe or Si substrates, clearly in which no Al diffusion occurs in the $Cr_2Te_3$ lattice (Figure S1). Since the major difference between the n-type and p-type $Cr_2Te_3$ lay in the Al incorporation caused by interfacial diffusion, hereby we would like to attribute the lattice constant dependence on thickness to the diffusion of Al atoms, which may cause the inflation of c-axis in our sample. However, such speculation deserves a systematic study by directing doping Al into the $Cr_2Te_3$ thin film in the future, which may also provide a controllable way for adjusting the curie temperature of $Cr_2Te_3$.

**CONCLUSIONS**

In this work, MBE-grown hexagonal $Cr_2Te_3$ thin films grown on $Al_2O_3$ substrates were found to have high Curie temperature. In-situ long streaky reflection high energy electron diffraction patterns observed in the $Cr_2Te_3$ thin films indicate that they were grown with atomically flat surface with high crystalline quality. It was found that a 2 nm amorphous layer of Al exists at the interface between $Cr_2Te_3$ and $Al_2O_3$ and a small amount of Al contamination is incorporated into the $Cr_2Te_3$ layer, which induces the n-type conduction type of our samples in contrast to the p-type signature observed for $Cr_2Te_3$ samples as reported by others previously. A phenomenological model for the coupling between FM ordering and AFM ordering that exists in our $Cr_2Te_3$ thin films is proposed based on the results of the magnetic characterizations performed on them. The phenomenon that Curie temperature increases as the c-lattice parameter increases can be explained using our proposed phenomenological model.

**ASSOCIATED CONTENT**

SUPPORTING INFORMATION

The stoichiometry analysis; HRTEM, R-T, Hall measurement of Contrast sample without amorphous layer.

FUNDING SOURCES

This work was supported by the National Natural Science Foundation of China (No. 11774143), the National Key Research and Development Program of China (No. 2018YFA0307100, No.2016YFA0301703), the Natural Science Foundation of Guangdong Province (No.2015A030313840, No.2017A030313033), the State Key Laboratory of Low-Dimensional Quantum Physics (No.KF201602), Technology and Innovation Commission of Shenzhen Municipality (KQJSCX20170727090712763, ZDSYS20170303165926217, and JCYJ20170412152334605).

NOTES

The authors declare no competing financial interest.